\documentclass[prc, reprint, amsmath, amssymb, superscriptaddress, nofootinbib, showpacs]{revtex4-1}

\usepackage{graphicx}%
\usepackage{multirow}
\usepackage{float}
\usepackage{epstopdf}
\usepackage[utf8]{inputenc}
\usepackage[english]{babel}
\usepackage{footnote}
\usepackage[justification=justified,
   format=plain]{caption}
\usepackage{xcolor}
\usepackage{textcomp}

\begin{document}

\title{Updated evaluation of potential ultra-low Q value $\beta$ decay candidates}%

\author{D. K. Keblbeck}
\affiliation{Department of Physics, Central Michigan University, Mount Pleasant, MI 48859, USA}

\author{R. Bhandari}
\affiliation{Department of Physics, Central Michigan University, Mount Pleasant, MI 48859, USA}

\author{N. D. Gamage}
\affiliation{Department of Physics, Central Michigan University, Mount Pleasant, MI 48859, USA}
\affiliation{Facility for Rare Isotope Beams, Michigan State University, East Lansing, Michigan, 48824 USA}

\author{M. Horana Gamage}
\affiliation{Department of Physics, Central Michigan University, Mount Pleasant, MI 48859, USA}

\author{K. G. Leach}
\affiliation{Department of Physics, Colorado School of Mines, Golden, Colorado 80401, USA}
\affiliation{Facility for Rare Isotope Beams, Michigan State University, East Lansing, Michigan, 48824 USA}

\author{X. Mougeot}
\affiliation{Universit{\'e} Paris-Saclay, CEA, List, Laboratoire National Henri Becquerel (LNE-LNHB), F-91120 Palaiseau, France}

\author{M. Redshaw}
\affiliation{Department of Physics, Central Michigan University, Mount Pleasant, MI 48859, USA}
\affiliation{National Superconducting Cyclotron Laboratory, East Lansing, Michigan, 48824 USA}

\date{\today}%

\begin{abstract}
``Ultra-low" Q value $\beta$ decays are referred to as such due to their low decay energies of less than $\approx$1 keV. Such a low energy decay is possible when the parent nucleus decays to an excited state in the daughter, with an energy close to that of the Q value. These decays are of interest as potential new candidates for neutrino mass determination experiments and as a testing ground for studies of atomic interference effects in the nuclear decay process. In this paper, we provide an updated evaluation of atomic mass data and nuclear energy level data to identify potential ultra-low Q value $\beta$ decay candidates. For many of these candidates, more precise and accurate atomic mass data is needed to determine if the Q value of the potential ultra-low decay branch is energetically allowed and in fact ultra-low. The relevant precise atomic mass measurements can be achieved via Penning trap mass spectrometry.
\end{abstract}

\maketitle

\section{Introduction}
From the proposed existence of the neutrino, to the development of electroweak theory, nuclear $\beta$-decay has played a central role in our understanding and development of sub atomic physics. Experimental measurements and theoretical descriptions of $\beta$-decay continue to impact nuclear and particle physics, for example, via experiments that place constraints on the neutrino mass~\cite{Aseev2011_3T,Kraus2005_3T,Aker2019,Arnaboldi2003,Springer1987_163Ho}, searches for sterile neutrinos~\cite{Friedrich_7Be,Martoff2021_HUNTER} and neutrinoless double $\beta$-decay~\cite{Avignone2008_0vbb,Blaum_0v2EC}, and via experiments that test the validity of the Standard Model (SM) and constrain physics beyond the SM~\cite{Falkowski2021,Gonzalez2019}.

In the context of direct neutrino mass determination experiments, the approach is to extract a value or upper limit for the (electron) neutrino mass from the slight distortion of the $\beta$-decay or electron capture (EC) de-excitation energy spectrum near the end-point due to a non-zero neutrino mass~\cite{Formaggio2021}. The fraction of decays in an energy interval $\Delta E$, near the end-point, goes as ($\Delta E/Q$)$^{3}$~\cite{Ferri2015} for $\beta$-decay, and as ($\Delta E/Q$)$^{2}$ for EC~\cite{Ge2021_159Dy} (but can be enhanced when the binding energy of the captured orbital electron is close to the Q value). Hence, a low Q value decay is an important characteristic for the isotope to be studied. $^{3}$H, $^{187}$Re, and $^{163}$Ho, have the lowest known ground-state to ground-state (gs-gs) Q values~\cite{Myers2015_3H,Nesterenko2014_187Re,Eliseev2015_163Ho} and have provided the most stringent upper limits on the neutrino and anti-neutrino masses to date. Nevertheless, even lower Q value decays are possible for certain isotopes when the decay occurs to an excited nuclear state in the daughter with an energy very close to the gs-gs Q value, $Q_{gs}$. Such an isotope could be of interest as a potential candidate for future direct neutrino mass determination experiments.

For $\beta^{-}$ and EC decay, $Q_{gs}$ is defined as the energy equivalent of the mass difference between parent ($P$) and daughter ($D$) atoms:
\begin{equation}
    Q_{gs}^{\beta^{-}} = [M_{P}(^{A}_{Z}P) - M_{D}(^{\quad A}_{Z+1}D)]c^{2},
    \label{Eqn_Qgs_b-}
\end{equation}
\begin{equation}
    Q_{gs}^{EC} = [M_{P^{'}}(^{A}_{Z}P^{'}) - M_{D^{'}}(^{\quad A}_{Z-1}D^{'})]c^{2},
    \label{Eqn_Qgs_EC}
\end{equation}
where $M_{P^{(')}}$ and $M_{D^{(')}}$ are the masses of the parent and daughter atoms, respectively. For $\beta^{+}$ decay, $Q_{gs}$ is reduced due to the two additional electrons whose mass needs to be accounted for:
\begin{equation}
    Q_{gs}^{\beta^{+}} = Q_{gs}^{EC} - 2m_{e}c^{2},
    \label{Eqn_Qgs_b+}
\end{equation}
where $m_{e}$ is the electron mass. The Q value for a decay to an excited state in the daughter with energy $E^{*}$ is given by 
\begin{equation}
    Q_{es} = Q_{gs} - E^{*},
    \label{Eqn_Qes}
\end{equation}
and could potentially be very low. A decay with $Q_{es} < $ 1 keV, has been dubbed as ``ultra-low''~\cite{Mustonen2010_ULQ}. 

The identification of the first, and currently only known ultra-low Q value decay was made by Cattadori \textit{et al.}~\cite{Cattadori2005_115In}. The authors inferred the existence of a weak and potentially very low Q value $\beta$-decay branch in $^{115}$In to the first excited state in $^{115}$Sn via the detection of a 497 keV $\gamma$-ray that was assumed to come from the subsequent decay of the $^{115}$Sn(3/2$^{+}$; 497.3 keV) state. A Q value of 2(4) keV was deduced using available mass data for $^{115}$In and $^{115}$Sn and the energy of the 497.334(22) keV 3/2$^{+}$ state in $^{115}$Sn~\cite{Blachot2005}. This decay was confirmed to be energetically allowed by independent Penning trap measurements of the $^{115}$In -- $^{115}$Sn mass difference performed at Florida State University~\cite{Mount2009_115In} and with JYLFTRAP at the University of Jyv\"askyl\"a~\cite{Wieslander2009_115In}. The resulting $Q_{es}$ values were 0.155(24) keV and 0.35(17) keV, respectively, indicating that the decay branch in question is energetically allowed, and making it the lowest-known Q value $\beta$-decay. Recent measurements of the $^{115}$In 3/2$^{+}$ daughter state energy provide results of 497.342(3) keV~\cite{Zhe2019} and 497.316(7) keV~\cite{Urban2015}. These results both agree with the previous value~\cite{Blachot2005} and reduce the uncertainty, but disagree with each other at the 3.4$\sigma$ level. 

In addition to their relevance to direct neutrino mass determination experiments, ultra-low Q value $\beta$-decays also serve as important test cases for probing the boundaries between nuclear and atomic physics and investigating atomic interference effects in nuclear $\beta$-decay. Theoretical studies of the ultra-low Q value decay branch of $^{115}$In~\cite{Mustonen2010_ULQ} revealed a discrepancy between the calculated and experimental half-life for the ultra-low decay branch. Theoretical calculations of the $^{115}$In gs-gs decay rate, however, were in reasonably good agreement with the experimental result, suggesting that the discrepancy may be due to atomic interference effects that become more significant when the Q value is very small. However, more cases are required for study in order to draw definite conclusions.

To push the sensitivity of $\beta$-decay-based neutrino mass measurements below the 0.1~eV sensitivity level, a suitable ultra-low Q candidate is required.  Given the extremely small phase-space for these decays, an experimentally suitable candidate needs to have a relatively large branching ratio and short half-life to the state of interest in the daughter (ie. a short partial half-life $t=\frac{T_{1/2}}{BR}$) to generate a signal in a reasonable experimental time. 

This paper is organized as follows: In section II, we review the literature on theoretical and experimental studies of potential ultra-low Q value $\beta$-decay candidates and discuss our updated evaluation of all potential ultra-low Q value decays across the nuclear chart. In section III we present our results for $\beta^{\pm}$ and EC decay candidates that could undergo an ultra-low Q value decay. In section IV we provide a discussion of our results and describe some of the experimental and theoretical challenges that would be involved in using an ultra-low Q value decay in a direct neutrino mass determination experiment. Section V then provides a brief conclusion.

\section{Evaluation of potential ultra-low Q value candidates}
After the discovery of the $^{115}$In ultra-low Q value $\beta$-decay branch, Suhonen, \textit{et al.} identified a number of other potential ultra-low Q value decay candidates, including $^{135}$Cs~\cite{Mustonen2011}, $^{115}$Cd~\cite{Haaranen2013}, and several others~\cite{Mustonen2010,Suhonen2014}. The use of very low Q value decays in the context of neutrino mass determination experiments was discussed earlier by Kopp and Merhl~\cite{Kopp2010}, who identified a number of potential candidates, and our previous work provided a large survey of potential ultra-low Q value candidates~\cite{Gamage2019}.

Since the publication of these previous works, a number of precise Penning trap mass measurements have been performed to investigate potential ultra-low Q value decay candidates. Significantly, the JYFLTRAP group confirmed that $^{135}$Cs does potentially have an energetically allowed and ultra-low Q value $\beta$-decay branch~\cite{deRoubin2020_135Cs} with $Q_{es}$ = 0.44(31) keV. They also determined that $^{159}$Dy~\cite{Ge2021_159Dy} and $^{111}$In~\cite{Ge2022_111In} have low energy EC decay branches (with $Q_{es} \sim$1 -- 4 keV) to excited states in their respective daughter nuclei, and that $^{131}$I has an energetically allowed $\beta$-decay branch with $Q_{es} \approx$ 1 keV~\cite{Eronen2022_131I}. All of these transitions could be of interest for future neutrino mass determination experiments. Measurements at JYFLTRAP~\cite{Ramalho_75As} and LEBIT~\cite{Horana2022_75As} also determined that $^{75}$Se has an energetically allowed EC decay branch to the 865.4(5) keV (3/2$^{-}$ or 5/2$^{-}$) state in $^{75}$As with $Q_{es} \lesssim$ 1 keV. However, a more precise determination of the 865.4 keV energy level is needed to further evaluate this decay channel.

Other potential ultra-low Q value decays branches in $^{115}$Cd, identified in~\cite{Haaranen2013}, and $^{75}$Ge, $^{89}$Sr, $^{139}$Ba, and $^{112,113}$Ag, identified in Ref.~\cite{Gamage2019}, have been investigated by the LEBIT group at the National Superconducting Cyclotron Facility~\cite{Sandler2019_89Y_139La} and the CPT group at Argonne National Laboratory~\cite{Gam2022_CPT} and have been ruled out as not being ultra-low. The ISOLTRAP group performed a precise measurement of the $^{131}$Cs EC Q value, indicating that it does not have an ultra-low Q value decay branch~\cite{Karthein2019_131Cs}, and JYLFTRAP recently investigated and ruled out potential ultra-low Q value decay branches in $^{72,76}$As, $^{75}$Ge, and $^{155}$Tb~\cite{Ze2020_72As,Ramalho_75As,Ge2022_76As_155Tb}.

\begin{figure}[b]
    \centering
    \includegraphics[width=1\columnwidth]{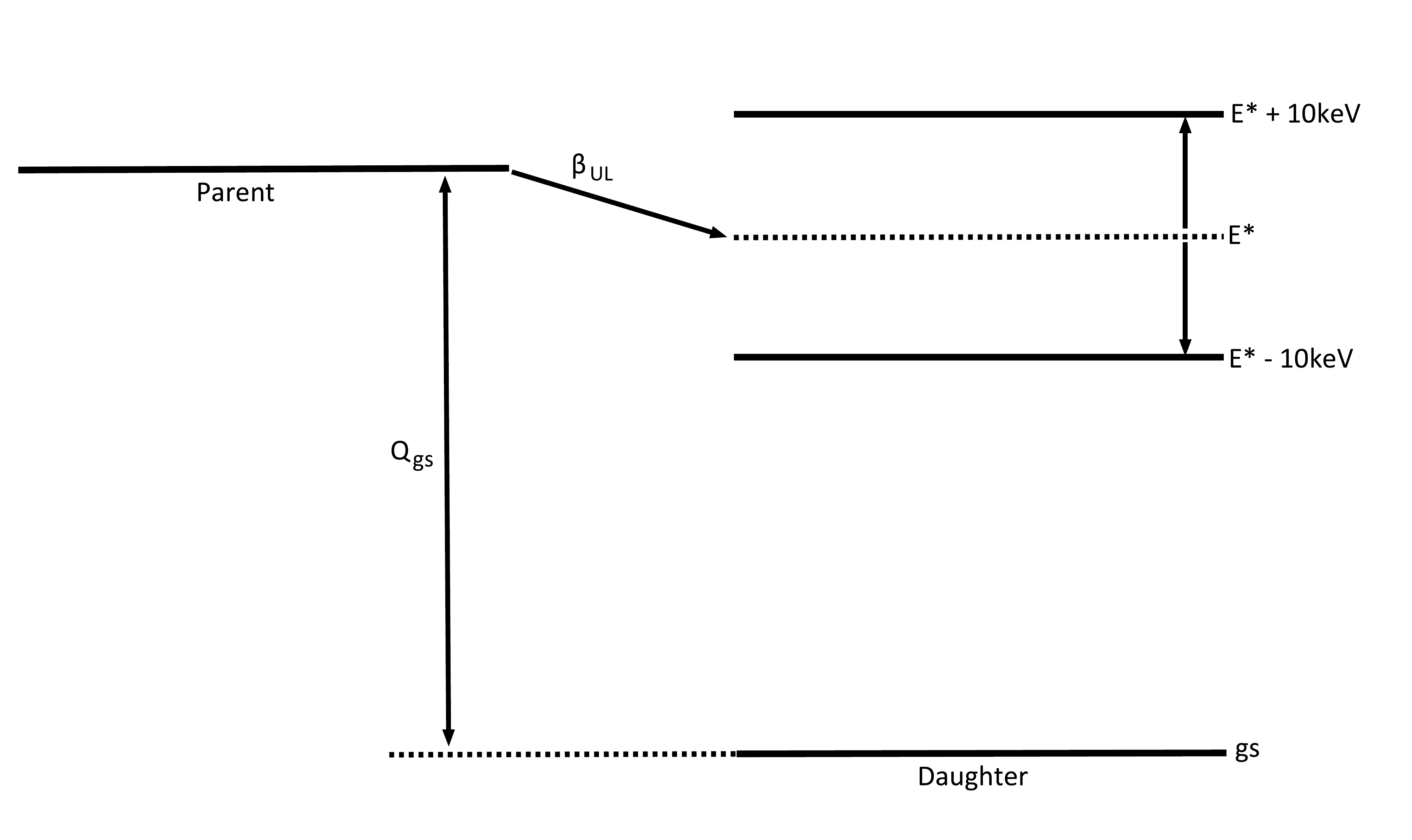}
    \caption{Decay scheme for potential ultra-low Q value candidates showing the energy range of the excited state of the daughter nucleus around the gs-gs Q value.}
    \label{fig:2}
\end{figure}

In this work, we provide an update to our previous survey of potential ultra-low Q value decay candidates using the recently published update to the Atomic Mass Evaluation (AME2020)~\cite{Wang_2021}, and most recent nuclear energy level data compiled by the National Nuclear Data Center~\cite{nndc}. In our evaluation we have identified candidates with a half-life $\geq$ 1 d, that could potentially undergo $\beta^{\pm}$ or EC decay to an excited state in the daughter nucleus. The main criteria we used to select candidates was that the difference between the gs-gs Q value and the excited state in the daughter nucleus be $\sim\pm$10 keV (see Fig. \ref{fig:2}). This range was chosen to account for the typical maximum size of observed shifts in atomic masses of isotopes close to stability after precise measurements with a Penning trap compared to values obtained from indirect methods see e.g. Refs.~\cite{Shi2005_KrXe,Redshaw2012_48Ca,Fink2012_110Pd}.

For this updated survey, we have taken into account binding energies of K, L, or M shell electrons for the EC decays, which modifies the Q value of an EC decay to an excited state in the daughter as follows:
\begin{equation}
    Q_{es}^{EC} = Q_{gs}^{EC} - E^{*} - E_{B},
    \label{Eqn_Qes_EC}
\end{equation}

where $E_{B}$ is the binding energy of the K, L, or M shell electron in the daughter, obtained from Ref.~\cite{Larkins1977}. For L and M shell capture, an average of the L1--L3 and M1--M5 energies, respectively, from Ref.~\cite{Larkins1977} were used $\footnote{The spread of the L1, L2, L3 and M1, M2, M3, M4, and M5 energies is much less than our criteria that $Q^{EC}_{es} - E_{B}$ be within $\pm$10 keV of $E^{*}$.}$.

The list of potential UL Q value candidates was further refined based on the uncertainties in $Q_{gs}$ and $E^{*}$: candidates with $\sigma(Q_{gs}) \gtrsim$ 10 keV and $\sigma(E^{*}) \gtrsim$ 1 keV were generally not included. Selection was also determined by whether or not the masses had been previously measured via Penning trap mass spectrometry. Decays with a forbiddenness greater than fourth-forbidden were also not considered since no such decays have been observed~\cite{Quarati2022_176Lu}.

\section{Results}
Our updated analysis provides a list of 77 potential ultra-low Q value $\beta$ and EC decay candidate isotopes, involving around 100 transitions to distinct energy levels in their respective daughter nuclei. These are listed in Tables \ref{UL_beta_table}, \ref{UL_EC_table} and \ref{UL_Beta+_table}. This analysis shows a decrease in the number of candidate isotopes compared to our previous evaluation~\cite{Gamage2019}, primarily due to the more stringent minimum half-life cutoff of one day that we imposed in this evaluation as compared to one hr in our previous one. This condition was chosen to provide a more realistic list of candidates that could be used for experimental study. Some additional candidates have been included in this evaluation due to changes in atomic masses between the 2016 and 2020 atomic mass evaluations, and due to the inclusion of electron binding energies in the $Q_{es}^{EC}$ calculation. The isotopes $^{72,76}$As, $^{75}$Ge, $^{89}$Sr, $^{112}$Ag, $^{115}$Cd, $^{131}$Cs, and $^{155}$Tb whose Q values have been precisely measured by Penning trap mass spectrometry in the last few years, ruling them out as not having ultra-low Q value transitions, have been omitted from these tables. $^{139}$Ba and $^{113}$Ag, that both do still have potential UL Q value decay transitions have also been left out of the current tables because they have half-lives of $<$ 1 d and the uncertainties in the energies of the relevant daughter states is $\approx$10 keV. Isotopes whose Q values have been measured with a Penning trap and have been shown to have a low or potentially ultra-low Q value decay branch e.g. $^{75}$Se, $^{111}$In,  $^{131}$I, $^{135}$Cs, and $^{159}$Dy have been included.

In the tables of results, Column I lists the parent isotope along with its ground state spin and parity assignment, and Column II its half-life. Column III lists the daughter isotope with spin and parity assignment for the excited state to which the potential ultra-low Q value decay would occur, and Column V the energy of that daughter state. Column IV provides the gs-gs Q value from Eqns. (\ref{Eqn_Qgs_b-}), (\ref{Eqn_Qgs_EC}) or (\ref{Eqn_Qgs_b+}) for $\beta^{-}$, EC, or $\beta^{+}$ decay, respectively. In Table \ref{UL_EC_table}, Column VII lists the electron binding energy for K, L, and M shell electrons (as listed in Column VI) in the daughter atom. Table \ref{UL_EC_table} is divided into three sections, upper for K shell capture, middle for L shell capture, and lower for M shell capture. The final two columns in the tables list the forbiddenness of the decay (see e.g. Ref. \cite{Suhonen_Nucleons} for details and \cite{ForbidCalc} for a forbiddenness calculator tool) and the Q value for a decay to the excited state, which can be considered as ultra-low if it is energetically allowed and $<$ 1 keV. The uncertainty in the ultra-low Q value includes the uncertainty from $Q_{gs}$ and $E^{*}$ (if available) added in quadrature. 


\begin{table*}[ht]
\centering
\caption{\label{UL_beta_table}Potential ultra-low Q value $\beta^{-}$ decays. Columns I and III list parent and daughter isotopes, respectively, and their spin and parity ($J^{\pi}$) assignments. $J^{\pi}$ assignments enclosed by braces indicates uncertain assignments, resulting in uncertainty in the decay type, indicated by a \{?\} in Column VI. For uncertain ($J^{\pi}$) assignments, we have listed the $J^{\pi}$ that results in the least-forbidden decay. The parent half-life is listed in Column II. Column IV lists the gs-gs Q value (Ref.~\cite{Wang_2021}); Column V, the excited state energy of the daughter nucleus (Ref.~\cite{nndc}) to which the potential ultra-low Q value decay would occur; and Column VII, the Q value of the decay to the excited state.}

\begin{ruledtabular}
\begin{tabular}{ccccccc} 
Parent & T$_{1/2}$ & Daughter & $Q_{gs}$ (keV) & $E^{*}$ (keV) & Decay type & $Q_{es}$ (keV)\\
\noalign{\smallskip}\hline\noalign{\smallskip}
$^{47}$Ca(7/2$^{-}$) & 5 d & $^{47}$Sc(3/2$^{+}$) & 1992.2(29) & 2002.30(30) & 1$^{\rm{st}}$ FU & -10.1(30) \\

$^{74}$As(2$^{-}$) & 18 d & $^{74}$Se(4$^+$) & 1353.1(17) & 1363.17(7) & 1$^{\rm{st}}$ FU & -10.0(17) \\

$^{77}$As(3/2$^{-}$) & 2 d & $^{77}$Se(5/2$^{+}$) & 683.2(17) & 680.10(20) & 1$^{\rm{st}}$ FNU & 3.1(17)\\

$^{98}$Tc\{6$^{+}$\} & 4.2 Myr & $^{98}$Ru(3$^{+}$) & 1792.7(73) & 1797.03(6) & 2$^{\rm{nd}}$ FU \{?\} & -4.4(73)\\

$^{124}$Sb(3$^{-}$) & 60 d & $^{124}$Te(2$^{+}$) & 2905.1(19) & 2897.3(10) & 1$^{\rm{st}}$ FNU & 7.8(22)\\

 & & $^{124}$Te\{7$^{-}$\} & 2905.1(19) & 2911.18(14) & 4$^{\rm{th}}$ FNU \{?\} & -6.1(19)\\

$^{127}$Sb(7/2$^{+}$) & 4 d & $^{127}$Te(5/2$^{+}$) & 1582.2(53) & 1568.13(11) & Allowed & 14.1(53)\\  

$^{131}$I(7/2$^{+}$) & 8 d & $^{131}$Xe\{9/2$^{+}$\} & 972.25(19) & 971.22(13) & Allowed \{?\} & 1.03(23)\footnote{Ref.~\cite{Eronen2022_131I}.}\\
        
$^{133}$Xe(3/2$^{+}$) & 5 d & $^{133}$Cs(1/2$^{+}$) & 427.4(24) & 437.01(1) & Allowed & -9.7(24)\\

$^{135}$Cs(7/2$^{+}$) & 2.3 Myr & $^{135}$Ba(11/2$^{-}$) & 268.66(30) & 268.218(20) & 1$^{\rm{st}}$ FU & 0.44(31)\footnote{Ref.~\cite{deRoubin2020_135Cs}.}\\

$^{136}$Cs(5$^{+}$) & 13 d & $^{136}$Ba(4$^{+}$) & 2548.2(19) & 2544.48(24) & Allowed & 3.7(19)\\

$^{140}$Ba(0$^{+}$) & 13 d & $^{140}$La(4$^{-}$) & 1044.2(79) & 1035.63(17) & 3$^{\rm{rd}}$ FU & 8.5(79)\\

 & & $^{140}$La(3$^{-}$) & 1044.2(79) & 1037.60(50) & 3${\rm{rd}}$ FNU & 6.6(79)\\

$^{140}$La(3$^{-}$) & 2 d & $^{140}$Ce\{3$^{+}$\} & 3762.2(14) & 3767.97(10) & 1$^{\rm{st}}$ FNU \{?\} & -5.8(14)\\

$^{148}$Pm(1$^{-}$) & 5 d & $^{148}$Sm\{3$^{-}$\} & 2470.2(58) & 2467.38(8) & 2$^{\rm{nd}}$ FNU \{?\} & 2.8(58)\\
              
$^{151}$Pm(5/2$^{+}$) & 1 d & $^{151}$Sm(5/2$^{+}$) & 1190.2(47) & 1188.0(20) & Allowed & 2.2(51)\\

$^{154}$Eu(3$^{-}$) & 9 yr & $^{154}$Gd\{2$^{+}$\} & 1968.0(15) & 1964.05(12) & 1$^{\rm{st}}$ FNU \{?\} & 3.9(16)\\

 & & $^{154}$Gd(2$^{+}$) & 1968.0(15) & 1973.07(17) & 1$^{\rm{st}}$ FNU  & -5.1(16)\\
              
$^{155}$Eu(5/2$^{+}$) & 5 yr & $^{155}$Gd(9/2$^{-}$) & 252.0(16) & 251.71(1) & 1$^{\rm{st}}$ FU & 0.3(16)\\

$^{156}$Eu(0$^{+}$) & 15 d & $^{156}$Gd\{2$^{+}$\} & 2452.5(37) & 2451.5$\footnote{An uncertainty for this energy level is not provided in Ref.~\cite{nndc}.}$ & 2$^{\rm{nd}}$ FNU \{?\} & 1.0(37)\\

 & & $^{156}$Gd(1$^{-}$) & 2452.5(37) & 2449.7$^{\textrm{c}}$ & 1$^{\rm{st}}$ FNU & 2.8(37)\\
 & & $^{156}$Gd(2$^{+}$) & 2452.5(37) & 2446.16(3) & 2$^{\rm{nd}}$ FNU & 6.3(37)\\
              
$^{161}$Tb(3/2$^{+}$) & 7 d & $^{161}$Dy(1/2$^{+}$) & 593.7(14) & 607.58(2) & Allowed & -13.9(14)\\

$^{166}$Dy(0$^{+}$) & 3 d & $^{166}$Ho(3$^{+}$) & 485.9(11) & 481.846(4) & 2$^{\rm{nd}}$ FU & 4.0(11)\\

$^{169}$Er(1/2$^{-}$) & 9 d & $^{169}$Tm(1/2$^{-}$) & 353.49(80) & 341.94(4) & Allowed & 11.55(80)\\

 &  & $^{169}$Tm(5/2$^{-}$) & 353.49(80) & 345.028(3) & 2$^{\rm{nd}}$ FNU & 8.46(80)\\

$^{171}$Tm(1/2$^{+}$) & 2 yr & $^{171}$Yb(7/2$^{+}$) & 96.55(97) & 95.282(2) & 2$^{\rm{nd}}$ FU & 1.26(97)\\

$^{182}$Ta(3$^{-}$) & 115 d & $^{182}$W(5$^{-}$) & 1815.5(17) & 1809.64(7) & 2$^{\rm{nd}}$ FNU & 5.8(17)\\
 & & $^{182}$W(6$^{-}$) & 1815.5(17) & 1810.85(4) & 2$^{\rm{nd}}$ FU & 4.6(17)\\ 
 
$^{183}$Ta(7/2$^{+}$) & 5 d & $^{183}$W(7/2$^{-}$) & 1072.1(18) & 1069.42(9) & 1$^{\rm{st}}$ FNU & 2.7(18)\\
 
$^{186}$Re(1$^{-}$) & 4 d & $^{186}$Os(4$^{+}$) & 1072.7(11) & 1070.48(3) & 3$^{\rm{rd}}$ FNU & 2.2(11)\\
 
$^{188}$W(0$^{+}$) & 70 d & $^{188}$Re\{1$^{+}$\} & 349.0(32) & 353.57(1) & Allowed \{?\} & -4.6(32) \\
 & & $^{188}$Re(4$^{+}$) & 349.0(32) & 342.59(2) & 4$^{\rm{th}}$ FNU & 6.4(32)\\

$^{189}$Re(5/2$^{+}$) & 1 d & $^{189}$Os(3/2$^{+}$) & 1007.7(82) & 996.40(40) & Allowed & 11.3(82)\\
           
$^{193}$Os(3/2$^{-}$) & 1 d & $^{193}$Ir\{9/2$^{-}$\} & 1141.9(27) & 1145.61(10) & 2$^{\rm{nd}}$ FU \{?\} & -3.7(27)\\

& &$^{193}$Ir(5/2$^{-}$) & 1141.9(27) & 1131.17(11) & Allowed & 10.7(27)\\
  
$^{199}$Au(3/2$^{+}$) & 3 d & $^{199}$Hg(3/2$^{-}$) & 452.31(76) & 455.46(2) & 1$^{\rm{st}}$ FNU & -3.15(76)\\ 
\end{tabular}
\end{ruledtabular}
\end{table*}


\begin{table*}
\centering
\caption{\label{UL_EC_table}Potential ultra-low Q value EC decays. See Table I for descriptions of Columns I -- V. Column VI lists the shell type. Column VII lists the binding energy of the K shell, L shell, or M shell electron of the daughter atom. Column VIII lists the decay type and Column IX the Q value of the decay to the excited state.}

\scalebox{1}{
\begin{ruledtabular}
\begin{tabular}{ccccccccc}

Parent & T$_{1/2}$ & Daughter & $Q_{gs}$ (keV) & $E^{*}$ (keV) & Shell & $E_{B}$ (keV) & Decay type & $Q_{es}$ (keV)\\

\noalign{\smallskip}\hline\noalign{\smallskip}

$^{56}$Co(4$^{+}$) & 78 d & $^{56}$Fe(4$^{+}$)& 4566.64(55) & 4554.77(9) & K & 7.1 & Allowed & 4.76(55)\\

$^{57}$Ni(3/2$^{-}$) & 1 d & $^{57}$Co\{3/2$^-$\} & 3261.70(77) & 3262.70(70) & K & 7.7 & Allowed \{?\} & -8.7(10)\\

$^{77}$Br(3/2$^{-}$) & 2 d & $^{77}$Se\{11/2$^{-}$\} & 1364.7(28) & 1351.58(12) & K & 12.7 & 4$^{\rm{th}}$ FNU \{?\} & 0.4(28)\\

$^{81}$Kr(7/2$^{+}$) & 229 kyr & $^{81}$Br(5/2$^{-}$) & 280.9(15) & 275.98(1) & K & 13.5 & 1$^{\rm{st}}$ FNU   & -8.6(15)\\

\noalign{\smallskip}\hline\noalign{\smallskip}

$^{56}$Co(4$^{+}$) & 78 d & $^{56}$Fe(4$^{+}$) & 4566.69(55) & 4554.77(9) & L & 0.8 & Allowed   & 11.12(55)\\

$^{57}$Ni(3/2$^{-}$) & 1 d & $^{57}$Co\{3/2$^-$\} & 3261.66(77) & 3262.70(70) &  L & 0.8 & Allowed \{?\} & -1.8(10)\\

 & & $^{57}$Co\{3/2$^-$\} & 3261.66(77) & 3272.2(11) & L & 0.8 & Allowed \{?\} & -11.3(13)\\

$^{73}$As(3/2$^{-}$) & 80 d & $^{73}$Ge\{5/2\} & 344.7(39) & 353.7(16) & L & 1.3 & Allowed \{?\} & -10.3(42)\\

$^{74}$As(2$^{-}$) & 18 d & $^{74}$Ge\{6$^{+}$\} & 2562.3(17) & 2569.33(14) & L & 1.3 & 3$^{\rm{rd}}$ FU \{?\}   & -8.3(17)\\
      
$^{77}$Br(3/2$^{-}$) & 2 d & $^{77}$Se\{3/2$^{-}$\} & 1364.7(28) & 1364.27(4) & L & 1.5 & Allowed \{?\} & -1.1(28)\\

$^{81}$Kr(7/2$^{+}$) & 229 kyr & $^{81}$Br(5/2$^{-}$)& 280.9(15) & 275.98(1) & L & 1.6 & 1$^{\rm{st}}$ FNU & 3.2(15)\\

$^{96}$Tc(7$^{+}$) & 4 d & $^{96}$Mo(5$^+$) & 2973.2(51) & 2975.28(7) & L & 2.7 & 2$^{\rm{nd}}$ FNU & -4.7(51)\\

 & & $^{96}$Mo(8$^{+}$) & 2973.2(51) & 2978.37(8) & L & 2.7 & Allowed & -7.8(51)\\
              
$^{97}$Tc(9/2$^{+}$) & 4.2 Myr & $^{97}$Mo(9/2$^{+}$) & 320.3(41) & 320.0(10) & L & 2.7 & Allowed & -2.4(42)\\

$^{101}$Rh(1/2$^{-}$) & 3 yr & $^{101}$Ru(7/2$^{+}$) & 545.7(59) & 545.11(1) & L & 3.0 & 3$^{\rm{rd}}$ FNU & -2.4(59)\\
              
$^{113}$Sn(1/2$^{+}$) & 115 d & $^{113}$In(1/2$^{+}$) & 1039.0(16) & 1029.65(5) & L & 4.0 & Allowed & 5.4(16)\\
              
$^{124}$I(2$^{-}$) & 4 d & $^{124}$Te(2$^{+}$) & 3159.6(27) & 3162.92(17) & L & 4.6 & 1$^{\rm{st}}$ FNU & -8.0(27)\\

$^{129}$Cs(1/2$^{+}$) & 1 d & $^{129}$Xe\{5/2$^{+}$\} & 1197.0(46) & 1197.11(21) & L & 5.1 & 2$^{\rm{nd}}$ FNU \{?\} & -5.2(46)\\

 & & $^{129}$Xe\{9/2$^{+}$\} & 1197.0(46) & 1194.60(30) & L & 5.1 & 4$^{\rm{th}}$ FNU \{?\} & -2.7(46)\\
             
$^{132}$Cs(2$^{+}$) & 7 d & $^{132}$Xe(6$^{+}$) & 2126.3(10) & 2111.88(16) & L & 5.1 & 4$^{\rm{th}}$ FNU & 9.3(10)\\

$^{146}$Pm(3$^{-}$) & 6 yr & $^{146}$Nd(2$^{+}$) & 1471.6(45) & 1470.63(6) & L & 6.7 & 1$^{\rm{st}}$ FNU & -5.8(45)\\

$^{147}$Eu(5/2$^{+}$) & 24 d & $^{147}$Sm(9/2$^{-}$) & 1721.4(29) & 1717.30(40) & L & 7.3 & 1$^{\rm{st}}$ FU & -3.1(29)\\

$^{156}$Tb(3$^{-}$) & 5 d & $^{156}$Gd(2$^{+}$) & 2444.3(39) & 2446.16(3) & L & 7.8 & 1$^{\rm{st}}$ FNU & -9.7(39)\\

 & & $^{156}$Gd\{2$^{+}$\} & 2444.3(39) & 2436.95(10) & L & 7.8 & 1$^{\rm{st}}$ FNU \{?\} & -0.5(39)\\
 
 & & $^{156}$Gd\{2$^{+}$\} & 2444.3(39) & 2434.70(1) & L & 7.8 & 1$^{\rm{st}}$ FNU \{?\}   & 1.8(39)\\

$^{157}$Tb(3/2$^{+}$) & 71 yr & $^{157}$Gd(5/2$^{-}$) & 60.0(14) & 54.54(1) & L & 7.8 & 1$^{\rm{st}}$ FNU & -2.3(14)\\

$^{167}$Tm(1/2$^{+}$) & 9 d & $^{167}$Er(7/2$^{-}$) & 746.1(13) & 745.32(10) & L & 9.1 & 3$^{\rm{rd}}$ FNU & -8.3(13)\\

$^{169}$Yb(7/2$^{+}$) & 32 d & $^{169}$Tm(11/2$^{-}$) & 899.12(76) & 884.62(20) & L & 9.5 & 1$^{\rm{st}}$ FU & 5.04(79)\\

$^{169}$Lu(7/2$^{+}$) & 1 d & $^{169}$Yb(5/2$^{+}$) & 2293.0(30) & 2286.2(12) & L & 9.8 & Allowed & -3.0(32)\\

 & & $^{169}$Yb(7/2$^{-}$) & 2293.0(30) & 2287.23(5) & L & 9.8 & 1$^{\rm{st}}$ FNU & -4.0(30)\\

$^{173}$Lu(7/2$^{+}$) & 1.5 yr & $^{173}$Yb(9/2$^{-}$) & 670.2(16) & 659.40(90) & L & 9.8 &  1$^{\rm{st}}$ FNU & 1.0(18)\\

$^{175}$Hf\{5/2$^{-}$\} & 70 d & $^{175}$Lu\{7/2$^{-}$\} & 683.9(26) & 672.83(15) & L & 10.2 & Allowed \{?\} & 1.0(26)\\
              
$^{177}$Ta(7/2$^{+}$) & 2 d & $^{177}$Hf\{11/2$^{-}$\} & 1166.0(36) & 1156.90(90) & L & 10.5 & 1$^{\rm{st}}$ FU \{?\} & -1.4(37)\\  

$^{183}$Re(5/2$^{+}$) & 70 d & $^{183}$W(9/2$^{-}$) & 556.0(81) & 551.24(3) & L & 11.3 & 1$^{\rm{st}}$ FU & -6.5(81)\\
   
$^{184}$Re\{3$^{-}$\} & 35 d & $^{184}$W(6$^{+}$) & 1485.6(43) & 1479.90(50)& L & 11.3 & 3$^{\rm{rd}}$ FNU \{?\} & -2.6(44)\\   

$^{190}$Ir(4$^{-}$) & 12 d & $^{190}$Os(2$^{+}$) & 1954.2(15) & 1943.50(40) & L & 12.1 & 1$^{\rm{st}}$ FU & -1.4(16)\\

$^{195}$Au(3/2$^{+}$) & 186 d & $^{195}$Pt(1/2$^{-}$) & 226.8(12) & 222.23(4) & L & 12.9 & 1$^{\rm{st}}$ FNU & -8.3(12)\\

$^{200}$Tl(2$^{-}$) & 1 d & $^{200}$Hg(1$^{-}$) & 2456.0(58) & 2442.70(30) & L & 13.8 & Allowed & -0.4(58)\\

\noalign{\smallskip}\hline\noalign{\smallskip}

$^{74}$As(2$^{-}$) & 18 d & $^{74}$Ge(4$^{+}$) & 2562.4(17) & 2569.33(14) & M & 0.1 & 1$^{\rm{st}}$ FU & -7.0(17)\\

$^{75}$Se(5/2$^{+}$) & 120 d & $^{75}$As\{5/2$^{-}$\} & 866.041(81) & 865.40(50) & M & 0.1 & 1$^{\rm{st}}$ FNU \{?\} & 0.54(51)\footnote{Q value from Ref.~\cite{Ramalho_75As}, see also Ref.~\cite{Horana2022_75As}.}\\

$^{77}$Br(3/2$^{-}$) & 2 d & $^{77}$Se\{3/2$^{-}$\} & 1364.7(28) & 1364.27(4) & M & 0.1 & Allowed \{?\} & 0.3(28)\\

$^{81}$Kr(7/2$^{+}$) & 229 kyr & $^{81}$Br(5/2$^{-}$) & 280.9(15) & 275.98(1) & M & 0.2 & 1$^{\rm{st}}$ FNU & 4.7(15)\\

$^{96}$Tc(7$^{+}$) & 4 d & $^{96}$Mo(2$^{+}$) & 2973.2(51) & 2975.28(7) & M & 0.4 & 4$^{\rm{th}}$ FU & -2.4(51)\\
& & $^{96}$Mo(2$^{+}$) & 2973.2(51) & 2978.37(8) & M & 0.4 & 4$^{\rm{th}}$ FU & -5.5(51)\\

\end{tabular}
\end{ruledtabular}}
\end{table*}

\begin{table*}
\ContinuedFloat
\centering
\caption{Continued.}

\scalebox{1}{
\begin{ruledtabular}
\begin{tabular}{ccccccccc} 

Parent & T$_{1/2}$ & Daughter & $Q_{gs}$ (keV) & $E^{*}$ (keV) & Shell & $E_{B}$ (keV) & Decay type & $Q_{es}$ (keV)\\

\noalign{\smallskip}\hline\noalign{\smallskip}

$^{97}$Tc(9/2$^{+}$) & 2.6 Myr & $^{97}$Mo(9/2$^{+}$) & 320.3(41) & 320.0(10) & M & 0.4 & Allowed & -0.1(42)\\

$^{101}$Rh(1/2$^{-}$) & 3 yr & $^{101}$Ru(7/2$^{+}$) & 545.7(59) & 545.11(1) & M & 0.4 & 3$^{\rm{rd}}$ FNU & 0.1(59)\\

$^{111}$In(9/2$^{+}$) & 3 d & $^{111}$Cd(7/2$^{+}$) & 857.63(17) & 853.94(7) & M & 0.6 & Allowed & 3.09(18)\footnote{Ref.~\cite{Ge2022_111In}.}\\
              
 & & $^{111}$Cd(3/2$^{+}$)& 857.63(17) & 855.6(10) & M & 0.6 & 2$^{\rm{nd}}$ FU & 1.4(10)$^{\textrm{a}}$\\

$^{113}$Sn(1/2$^{+}$) & 115 d & $^{113}$In(1/2$^{+}$) & 1039.0(16) & 1029.65(5) & M & 0.6 & Allowed & 8.7(16)\\

$^{124}$I(2$^{-}$) & 4 d & $^{124}$Te(2$^{+}$) & 3159.6(27) & 3149.50(70) & M & 0.8 & 1$^{\rm{st}}$ FNU & 9.3(28)\\
& & $^{124}$Te(2$^{+}$) & 3159.6(27) & 3154.37(3) & M & 0.8 & 1$^{\rm{st}}$ FNU & 4.4(27)\\
& & $^{124}$Te(2$^{+}$) & 3159.6(27) & 3162.92(17) & M & 0.8 & 1$^{\rm{st}}$ FNU & -4.1(27)\\
& & $^{124}$Te(2$^{+}$) & 3159.6(27) & 3167.94(8) & M & 0.8 & 1$^{\rm{st}}$ FNU & -9.1(27)\\

$^{129}$Cs(1/2$^{+}$) & 1 d & $^{129}$Xe\{5/2$^{+}$\} & 1197.0(46) & 1194.60(30) & M & 0.9 & 2$^{\rm{nd}}$ FNU \{?\} & 1.5(46)\\
& & $^{129}$Xe\{5/2$^{+}$\} & 1197.0(46) & 1197.11(21) & M & 0.9 & 2$^{\rm{nd}}$ FNU \{?\} & -1.0(46)\\

$^{146}$Pm(3$^{-}$) & 5 yr & $^{146}$Nd(2$^{+}$) & 1471.6.6(46) & 1470.63(6) & M & 1.3 & 1$^{\rm{st}}$ FNU & -0.3(45)\\

$^{146}$Eu(4$^{-}$) & 5 d & $^{146}$Sm(0$^{+}$) & 3878.8(67) & 3869.7(10) & M & 1.4 & 3$^{\rm{rd}}$ FU & 7.7(68)\\

$^{156}$Tb(3$^{-}$) & 5 d & $^{156}$Gd\{2$^{+}$\} & 2444.3(39) & 2436.95(10) & M & 1.5 & 1$^{\rm{st}}$ FNU \{?\} & 5.9(39)\\
& & $^{156}$Gd\{2$^{+}$\} & 2444.3(39) & 2442.41(10) & M & 1.5 & 1$^{\rm{st}}$ FNU \{?\} & 0.4(39)\\
& & $^{156}$Gd\{2$^{+}$\} & 2444.3(39) & 2446.16(3) & M & 1.5 & 1$^{\rm{st}}$ FNU \{?\} & -3.3(39)\\

$^{159}$Dy(3/2$^{-}$) & 144 d & $^{159}$Tb(5/2$^{-}$) & 364.73(19) & 363.545(1) & M & 1.6 & Allowed & -0.42(19)\footnote{Note that the transition to the $^{159}$Tb(5/2$^{-})$ state is energetically allowed for N and O shell capture, see Ref.~\cite{Ge2021_159Dy}.}\\

 & & $^{159}$Tb(11/2$^{+}$) & 364.73(19) & 362.05(4) & M & 1.6 & 3$^{\rm{rd}}$ FU & 1.08(19)$^{\textrm{b}}$\\

$^{167}$Tm(1/2$^{+}$) & 9 d & $^{167}$Er\{3/2$^{-}$\} & 746.1(13) & 745.32(10) & M & 1.8 & 1$^{\rm{st}}$ FNU \{?\} & -1.0(13)\\
& & $^{167}$Er\{3/2$^{-}$\} & 746.1(13) & 752.72(8) & M & 1.8 & 1$^{\rm{st}}$ FNU \{?\} & -8.4(13)\\

$^{173}$Lu(7/2$^{+}$) & 1.4 yr & $^{173}$Yb(9/2$^{-}$) & 670.2(16) & 659.40(90) & M & 1.9 & 1$^{\rm{st}}$ FNU & 8.9(18)\\

$^{174}$Lu\{1$^{-}$\} & 3.3 yr & $^{174}$Yb(3$^{-}$) & 1374.23(16) & 1382.01(6) & M & 1.9 & 2$^{\rm{nd}}$ FNU \{?\} & -9.7(16)\\

$^{177}$Ta(7/2$^{+}$) & 2 d & $^{177}$Hf\{11/2$^{-}$\} & 1166(36) & 1156.90(90) & M & 2.1 & 1$^{\rm{st}}$ FU \{?\} & 7.0(37)\\

$^{183}$Re(5/2$^{+}$) & 70 d & $^{183}$W\{5/2$^{-}$\} & 556.0(81) & 551.24(3) & M & 2.3 & 1$^{\rm{st}}$ FNU \{?\} & 2.5(81)\\
& & $^{183}$W\{5/2$^{-}$\} & 556.0(81) & 558.50(70) & M & 2.3 & 1$^{\rm{st}}$ FNU \{?\} & -3.8(81)\\

$^{184}$Re\{3$^{-}$\} & 38 d & $^{184}$W\{5$^{-}$\} & 1485.6(43) & 1476.90(50) & M & 2.3 & 2$^{\rm{st}}$ FNU \{?\} & 6.5(44)\\

$^{190}$Ir(4$^{-}$) & 12 d & $^{190}$Os\{2$^{+}$\} & 1954.2(15) & 1943.50(40) & M & 2.5 & 1$^{\rm{st}}$ FU \{?\} & 8.3(16)\\
& & $^{190}$Os\{2$^{+}$\} & 1954.2(15) & 1956.60(40) & M & 2.5 & 1$^{\rm{st}}$ FU \{?\} & -4.8(16)\\
& & $^{190}$Os\{2$^{+}$\} & 1954.2(15) & 1958.10(30) & M & 2.5 & 1$^{\rm{st}}$ FU \{?\} & -6.3(15)\\

$^{194}$Hg(0$^{+}$) & 440 yr & $^{194}$Au\{2$^{-}$\} & 28.0(36) & 35.19(7) & M & 2.8 & 1$^{\rm{st}}$ FU \{?\} & -10.0(36)\\

$^{195}$Au(3/2$^{+}$) & 186 d & $^{195}$Pt(5/2$^{-}$) & 226.8(12) & 222.23(4) & M & 2.7 & 1$^{\rm{st}}$ FNU & 1.9(12)\\

$^{200}$Tl(2$^{-}$) & 1 d & $^{200}$Hg\{1$^{+}$\} & 2456.0(58) & 2461.83(4) & M & 2.9 & 1$^{\rm{st}}$ FNU \{?\} & -8.7(58)\\

$^{205}$Bi(9/2$^{-}$) & 15 d & $^{205}$Pb(9/2$^{+}$) & 2704.6(49) & 2707.72(11) & M & 3.1 & 1$^{\rm{st}}$ FNU & -6.2(49)\\

\end{tabular}
\end{ruledtabular}}
\end{table*}


\begin{table*}
\centering
\caption{\label{UL_Beta+_table}Potential ultra-low Q value $\beta^+$ decays. See Table I for descriptions of Columns I -- VII.}

\makebox[\textwidth]{
\begin{ruledtabular}
\begin{tabular}{ccccccc} 
Parent & T$_{1/2}$ & Daughter & $Q_{gs}$ (keV) & $E^{*}$ (keV) & Decay type & $Q_{es}$ (keV)\\
\noalign{\smallskip}\hline\noalign{\smallskip}

$^{48}$V(4$^{+}$) & 16 d & $^{48}$Ti(0$^+$) & 2992.95(97) & 2997.22(16) & 4$^{\rm{th}}$ FNU & -4.27(99)\\

$^{56}$Ni(0$^{+}$) & 6 d & $^{56}$Co(3$^+$) & 1110.87(62) & 1114.51(5) & 2$^{\rm{nd}}$ FU & -3.64(62)\\

$^{79}$Kr(1/2$^{-}$) & 1 d & $^{79}$Br(3/2$^-$) & 603.8(36) & 606.03(6) & Allowed      & -2.2(36)\\

$^{83}$Sr(7/2$^{+}$) & 1 d & $^{83}$Rb\{5/2$^+$\} & 1251.0(72) & 1242.90(40) & Allowed \{?\} & 8.1(72)\\

$^{105}$Ag(1/2$^{-}$) & 41 d & $^{105}$Pd(5/2$^+$) & 325.1(47) & 319.38(22) & 1$^{\rm{st}}$ FU & 5.7(47)\\

$^{144}$Pm(5$^{-}$) & 1 yr & $^{144}$Nd(4$^+$) & 1309.9(32) & 1314.67(13) & 1$^{\rm{st}}$ FNU & -4.8(32)\\

$^{145}$Eu(5/2$^{+}$) & 6 d & $^{145}$Sm(3/2$^+$) & 1637.9(34) & 1627.74(4) & Allowed & 10.1(34)\\

$^{146}$Pm(3$^{-}$) & 6 yr & $^{146}$Nd(2$^+$) & 449.6(45) & 453.84(3) & 1$^{\rm{st}}$ FNU & -4.3(45)\\

$^{146}$Eu(4$^{-}$) & 5 d & $^{146}$Sm(2$^+$) & 2856.8(67) & 2859.0(10) & 1$^{\rm{st}}$ FU & -2.2(68)\\

$^{148}$Eu(5$^{-}$) & 55 d & $^{148}$Sm(4$^-$) & 2016.6(100) & 2031.4(13) & Allowed & -14.8(101)\\

$^{153}$Tb(5/2$^{+}$) & 2 d & $^{153}$Gd(5/2$^-$) & 547.3(41) & 548.77(2) & 1$^{\rm{st}}$ FNU  & -1.4(41)\\

 & & $^{153}$Gd\{5/2$^{-}$\} & 547.3(41) & 551.09(2) & 1$^{\rm{st}}$ FNU \{?\} & -3.8(41)\\

$^{169}$Lu(7/2$^{+}$) & 1 d & $^{169}$Yb\{7/2$^-$\} & 1271.0(30) & 1261.89(17) & 1$^{\rm{st}}$ FNU \{?\} & 9.1(30)\\

 &  & $^{169}$Yb\{1/2$^-$\} & 1271.0(30) & 1270.74(8)  & 3$^{\rm{rd}}$ FNU \{?\} & 0.3(30)\\

& &  $^{169}$Yb\{7/2$^+$\} & 1271.0(30) & 1285.13(8) & Allowed \{?\} & -14.1(30)\\

$^{171}$Lu(7/2$^{+}$) & 8 d & $^{171}$Yb(11/2$^-$) & 456.4(19) & 449.60(2) & 1$^{\rm{st}}$ FU & 6.8(19)\\

$^{188}$Ir(1$^{-}$) & 2 d & $^{188}$Os(0$^+$) & 1770.3(95) & 1765.40(40) & 1$^{\rm{st}}$ FNU & 4.9(95)\\

\end{tabular}
\end{ruledtabular}}
\end{table*}


\section{Discussion}

\subsection{Verification or elimination of potential candidates}
The goal of identifying additional ultra-low Q value $\beta$-decay candidates requires more precise $Q_{es}$ value determinations for many of the isotopes listed in Table I -- III. In most cases this will involve more precise measurements of the gs-gs Q value, $Q_{gs}$, determined from the mass difference between parent and daughter isotopes. This quantity can be accurately and precisely determined using Penning trap mass spectrometry, see e.g. Ref.~\cite{Brown1986,Blaum2006}. In some cases, the uncertainty in the energy of the final state in the daughter isotope may also need to be determined more precisely, particularly if it is found that $Q_{es}$ $<$ 1 keV, with an uncertainty dominated by the uncertainty in $E^{*}$. Although all of the candidates listed in these tables are of potential interest, the allowed, first-forbidden unique, and first-forbidden non-unique decays are of most interest for neutrino mass determination experiments and for theoretical and experimental comparison for studies of atomic interference effects. As such, we suggest a short list of candidates from these tables to be prioritized in future studies. These are allowed and first-forbidden decays with $Q_{es}$ values typically within a few keV of zero and with correspondingly small uncertainties in $Q_{es}$ that are dominated by the uncertainty in $Q_{gs}$. These are listed in Table \ref{UL_ShortList_Table}. We note that the evaluation of partial half-lives for these decays would require dedicated calculations to be performed on a case-by-case basis, as is discussed in more detail in section IV. C.

\subsection{Prospects for decay measurements with UL Q value candidates}

The primary challenge for experiment in using these UL Q value cases to perform neutrino mass measurements is the very long partial half-lives, $t=\frac{T_{1/2}}{BR}$, for decay to the state(s) of interest in the daughter. The observed $^{115}$In(9/2$^{+}$, gs) $\rightarrow$ $^{115}$Sn(3/2$^{+}$) UL Q value decay, for example, has a branching ratio of $\sim$10$^{-6}$~\cite{Wieslander2009_115In} and a parent half-life of $>10^{14}$ years which yields a partial half-life to that state of $t\geq10^{20}$ years -- roughly on the same order as double $\beta$-decay.  These issues can be addressed by a careful choice of the UL Q value candidate. To maximize the branching ratio and decay rate and simplify the analysis of the $\beta$-spectrum, an allowed or first forbidden transition is strongly preferred. An UL Q value decay to an excited daughter state that is not an isomer and has a single $\gamma$-transition or simple multiple $\gamma$-cascade to the ground state would provide also a convenient $\gamma$-ray or $\gamma$-rays to gate on to identify the $\beta$-decay signal. Finally, a parent isotope with a sufficiently long total half-life, $T_{1/2}$, would be required to allow for isotope harvesting to enable a controlled, off-line experiment with sufficient statistics to be performed. A large fraction of the isotopes listed in Table \ref{UL_ShortList_Table} will be able to be produced at the Facility for Rare Isotope Beams (FRIB) with rates of $\sim$10$^{7}$ -- 10$^{9}$ particles/s $\footnote{The quoted rates are stopped beam rates, so that ions could be implanted into a detector, for example. They also represent the ``ultimate FRIB yields''~\cite{FRIBRates}.}$, and for some cases even higher at ISOL-type facilities such as TRIUMF and CERN-ISOLDE. Hence, one day of isotope accumulation at FRIB could provide $\sim$10$^{12}$ -- 10$^{14}$ parent nuclides. Even with partial half-lives of $t\geq10^{14}$~s or longer, this would allow for enough statistics to perform a high-precision $\beta$ or EC spectrum measurement given a sufficient running period, and multiple loading of the radioactive sample into the measurement setup.

Assuming a suitable UL Q value candidate is identified, among the chief obstacles of such a case for neutrino mass measurements is the tremendous challenge it provides for experiment, both on the sub-keV energy-scale sensitivity needed and the precision required for a measurement of the effective neutrino mass below the $\sim0.1$~eV/$c^2$ level.  Recent advancements in low-energy quantum sensing have opened sensitive avenues for performing neutrino mass studies using the kinetic energy of the nuclear recoil and atomic relaxation following EC decay (c.f. Refs.~\cite{Gastaldo2017,Alpert2015,Friedrich_7Be}). These methods can also be extended to precision measurements of nuclear $\beta^\pm$ decay.

The main experimental criteria for exploiting UL Q value decays with such energy resolved detectors is sub-keV sensitivity to the endpoint of the decay radiation with sub-eV energy resolution. Not only is this challenging from a detector standpoint, it also requires complete knowledge of molecular and atomic broadening effects of the initial and final states -- something that poses a number of challenges~\cite{Fretwell2020,Samanta2022}.

Direct momentum measurements of the decay products, however, have significantly less strict experimental requirements than those that use energy resolving sensors since the momentum carried by secondary atomic particles emitted following the decay is much smaller than the nuclear recoil momentum.  Such measurements also avoid the atomic smearing that is inherent in the nuclear recoil energy discussed above.  In this respect, UL Q value cases are very attractive, and large increases in sensitivity to the neutrino mass can be achieved by searching for eV-level Q values that are experimentally accessible.  Although challenging, recent work has suggested that such measurements could be made using optically levitated nanospheres as mechanical quantum sensors~\cite{Carney2022}.  In fact, the authors of Ref.~\cite{Carney2022} explicitly state that,
\begin{quote}
    Realizing an optomechanical measurement of the light $\nu$ masses would likely require a transition with sufficiently low Q value, experimentally manageable $T_{1/2}$, and high branching ratio to be identified through such measurements. While speculative, the existence of such a transition is plausible, and we strongly encourage the precision atomic mass measurement community to continue their work in this area.
\end{quote}

As these techniques mature, the search for suitable UL Q value decays with the criteria outlined above becomes increasingly important, and sets the bar for the evaluation presented here.

\subsection{Challenges for theoretical calculations}

\begin{table}[t]
\centering
\caption{\label{UL_ShortList_Table} List of some of the most promising allowed and first forbidden potential ultra-low Q value $\beta^{\pm}$ and EC decay candidates to investigate. Candidates were selected from Tables I -- III for having a relatively long total half-life---making an experimental search for such a decay more feasible, and a $Q_{es}$ value around zero with relatively small uncertainty---making it more likely that the $Q_{es}$ value is in fact $\lesssim$ 1 keV.}

\begin{ruledtabular}
\begin{tabular}{ccccc} 
Isotope & Decay & Forbiddenness & Half-life & $Q_{es}$ (keV)\\
\hline
$^{136}$Cs & $\beta^{-}$ & Allowed & 13 d & 3.7(19)\\
$^{188}$W & $\beta^{-}$ & Allowed & 70 d & -4.6(32)\\
$^{155}$Eu & $\beta^{-}$ & 1$^{\textrm{st}}$ Forbidden & 5 yr & 0.3(16)\\
$^{156}$Eu & $\beta^{-}$ &1$^{\textrm{st}}$ Forbidden & 15 d & 1.0(37)\\
\hline
$^{56}$Co & EC & Allowed & 78 d & 4.76(55)\\
$^{97}$Tc & EC & Allowed & 4.2 Myr & -0.1(42)\\
$^{175}$Hf & EC & Allowed & 70 d & 1.0(26)\\
$^{81}$Kr & EC & 1$^{\textrm{st}}$ Forbidden & 229 kyr & 3.2(15)\\
$^{146}$Pm & EC &1$^{\textrm{st}}$ Forbidden & 6 yr & -0.3(45)\\
$^{157}$Tb & EC &1$^{\textrm{st}}$ Forbidden & 71 yr & -2.3(14)\\
$^{173}$Lu & EC &1$^{\textrm{st}}$ Forbidden & 1.5 yr & 1.0(18)\\
$^{183}$Re & EC &1$^{\textrm{st}}$ Forbidden & 70 d & 2.5(81)\\
$^{195}$Au & EC &1$^{\textrm{st}}$ Forbidden & 186 d & 1.9(12)\\
\hline
$^{148}$Eu & $\beta^{+}$ & Allowed & 55 d & -15(10)\\
$^{105}$Ag & $\beta^{+}$ & 1$^{\textrm{st}}$ Forbidden & 41 d & 5.7(47)\\
$^{144}$Pm & $\beta^{+}$ &1$^{\textrm{st}}$ Forbidden & 1 yr & -4.8(32)\\
$^{146}$Pm & $\beta^{+}$ &1$^{\textrm{st}}$ Forbidden & 6 yr & -4.3(45)\\
\end{tabular}
\end{ruledtabular}
\end{table}

Theoretical predictions of allowed $\beta$-decays are nowadays achievable at the 0.1\% precision level in the vast majority of the cases \cite{Hayen2018}. However, the calculation of a $\beta$-spectrum with an ultra-low end-point energy, and of the related integrated quantities like the $f$-value, the decay constant, the partial half-life and the branching ratio, still remains extremely challenging. The usual approximations reach their limits and every correction that is generally neglected, or assumed to be sufficiently well-known, can have a major influence \cite{Mustonen2010_ULQ}.

The most complete $\beta$-decay formalism is probably from Behrens and B{\"u}hring \cite{Behrens82}, in which the Hamiltonian density is defined from the nuclear and lepton currents. The core of the method is a double multipole expansion of these currents. Depending on the change of angular momentum and parity between the initial and final nuclear states, rules emerge for the selection of the terms that dominantly contribute to the transition probability. In addition, the lepton wave functions are expanded in powers of electron mass and energy, nuclear radius, fine structure constant and atomic number, i.e. ($m_e R$), ($WR$) and ($\alpha Z$), and only the first orders are kept. These usual prescriptions have been shown not to be sufficient to accurately describe $\beta$-decay observables hindered because of large angular momentum change or low transition energy ($\lesssim$ 100 keV), or both (see references in \cite{Haaranen2017}). The extension to next-to-leading order terms in the lepton current has been studied in \cite{Haaranen2017}. However, the method is still based on the assumption that $\beta$-particles feel only the Coulomb potential of an extended nucleus. Taking into account a more elaborated potential for the screening effect due to the atomic electrons prevents any expansion of the lepton wave functions. A possibility can be a full numerical calculation of the lepton current, thus virtually including all terms of the expansion. The computational effort is much more significant and has recently been investigated in the $^{151}$Sm low-energy transitions \cite{Kossert2022_151Sm}.

Atomic corrections have been demonstrated to have a significant influence on the $\beta$-spectrum shape, especially at very low energies ($\lesssim$ 10 keV), both directly \cite{Mougeot2012, Loidl2014, Mougeot2014, Hasel2020} and indirectly \cite{Kossert2015, Kossert2018, Kossert2021}. Screening has already been discussed above. Another correction is due to the overlap effect that comes from the mismatch of the initial and final atomic wave functions due to the sudden change in the number of protons, and that thus accounts for shake-up and shake-off effects. This effect is usually negligible but should play a more important role in transitions with ultra-low end-point energy as it tends to reduce the end-point energy by tens to hundreds of eV. The current available correction is only first order and depends on binding energies from an approximate fit \cite{Hayen2018}. It should be revised as its precision is questionable in the present context. Most importantly, the exchange effect has the strongest impact on the spectrum shape and the emission probabilities at very-low energies. Precise modeling exists for allowed transitions and has recently been extended to forbidden unique transitions \cite{Hasel2020}. At ultra-low energies, the precision of the formalism used to generate the atomic wave functions becomes of high importance. Deep studies would be necessary with a comparison of different many-body approaches and different electron correlation treatments in order to quantify their influence on the $\beta$-decay observables. In addition, the subtle interplay between the atomic and nuclear matrix elements is neglected but might have an influence in the case of forbidden non-unique transitions.

An important challenge comes obviously from the accuracy of the description of the nuclear structure and of its inclusion in the $\beta$-decay formalism. The latter must be totally relativistic due to the small rest mass of the $\beta$ particle. This implies the existence of relativistic transition matrix elements that couple small and large components of the nucleon wave functions. Such matrix elements are strictly speaking null with nuclear wave functions from the most common non-relativistic formalism, while they can be dominant e.g. in forbidden non-unique transitions. A long-standing approach is to estimate them by applying either the conserved vector current (CVC) hypothesis for the vector matrix elements, or the partially conserved axial-vector current (PCAC) hypothesis for the axial-vector matrix elements \cite{Behrens82}. The former depends on the Coulomb displacement energy between the initial and final nuclear states, for which it is not always easy to get an accurate estimate as it can be sensitive to the mismatch of the nucleon wave functions \cite{Damgaard1966}. Another effect is weak magnetism, which emerges as an induced interaction when considering nucleons of finite size. In the context of ultra-low end-point energies, its accuracy should not be a constraint as it appears to first order as a correction on the spectrum shape linear in energy (see e.g. in \cite{Hayen2018}). However, it has never been studied in detail in the case of forbidden transitions.

Regarding nuclear structure, different approaches exist depending on the mass region, such as the realistic shell model with fitted Hamiltonian from the NushellX code \cite{Brown2014} or the microscopic quasiparticle-phonon model \cite{Toivanen1998}. Nucleus deformation can also be incorporated \cite{Martini2016}. It has been shown that the spectrum shape can be strongly influenced by the value of the axial-vector coupling constant $g_A$ \cite{Kos17_1, Kos17_2}. In fact, an effective renormalization of $g_A$ may be necessary to account for nuclear medium effects and limitations of the many-body treatment, the constant appearing usually as quenched compared to its free-nucleon value. As an example, one can mention core polarization effects due to the restricted nucleon valence space employed \cite{Warb90, Warb91}. Thanks to the excellent review from Suhonen \cite{Suhonen2017}, an effective $g_A$ value can be estimated from the relationship given between infinite nuclear matter and finite nuclei. However, one cannot expect nowadays a high-precision predictive power and it is still necessary to have a case-by-case approach comparing theory and experiment. On the other hand, tremendous progress has been made in the past decade in nuclear theory, in particular by bridging the gap between low-energy quantum chromodynamics and microscopic nuclear forces \cite{Epelbaum2009}. It has led to the framework of the chiral effective field theory that considers nucleons and pions as effective degrees of freedom instead of quarks and gluons, and in which many-body nucleon forces naturally emerge \cite{Machleidt2011}. Applied with ab initio many-body methods, high-precision nuclear structure calculations become possible, controlling the precision and quantifying the theoretical uncertainties \cite{Carlsson2016}. Such an approach has been applied successfully to the determination of transition rates of allowed $\beta$-decays from $^{3}$H to $^{100}$Sn \cite{Gysbers2019}, and should make useless any effective $g_A$ constant. 

The formalism of Behrens and B{\"u}hring is a low-energy effective theory, not renormalizable, and only includes the electrostatic part of the Coulomb interaction that influences the $\beta$-decay observables. Non-static Coulomb corrections, called radiative corrections, have to be determined in the framework of quantum electrodynamics. Historically, they have been split into inner and outer corrections. The former can be calculated with high-precision and typically appears as a renormalization of the coupling constants. The latter have an influence on the $\beta$-spectrum shape with a dependency on both the lepton and nuclear terms. Analytical high-precision formulation has been established over the years, including the influence of nuclear structure, but only in the case of (super-)allowed decays \cite{Towner2008, Sirlin2011}. There is still a need for a detailed theoretical study in the case of the forbidden $\beta$ transitions. In addition, the outer corrections include the effect of internal bremsstrahlung in which a real photon is created in the final state, reducing the kinetic energy of the $\beta$ particle. If photons can be emitted up to the transition energy, the process is much more probable at low energy. These photons can be (partially) re-absorbed in the detection system, depending on the actual configuration. High-precision measurements of ultra-low end-point energy spectra would thus require a careful experimental analysis, for which a precise knowledge of the photon spectrum is needed.

To summarize, several challenges subsist in atomic physics, nuclear physics and their inclusion in the weak interaction decay formalism for a high-precision description of the $\beta$-spectrum and integrated quantities in the case of ultra-low transition energies. Different high-level expertise would have to be gathered to tackle this problem.

 \section{Conclusions}

In summary, so-called ultra-low Q value $\beta$-decays (with $Q$ $<$ 1 keV) to excited nuclear states in the daughter isotope present interesting possibilities for future direct neutrino mass determination experiments and also to test and develop theoretical understanding of the role of atomic interference effects in nuclear $\beta$-decay. Over the last decade or so, interest in these decays has grown, and the $Q_{gs}$ values for 16 candidates have so far been investigated via Penning trap mass spectrometry at five facilities world-wide. In addition to the verification of the ultra-low Q value decay of $^{115}$In, four of these additional candidates have been shown to have energetically allowed decay branches to excited nuclear states in the daughter with $Q_{es}$ values of around 1 keV or less, indicating the existence of new potential ultra-low Q value decay candidates. Other potential decays have been ruled out. Our updated evaluation lists $\approx$80 isotopes that could potentially have ultra-low Q value decay branches, but whose gs-gs Q values and in some cases excited daughter state energies need to be determined more precisely. The Q values can be readily determined to the necessary precision via precise mass measurements of the parent and/or daughter isotopes using Penning trap mass spectrometry. A short list of these candidates that are of particular interest to neutrino physics
and nuclear $\beta$-decay studies due to the fact that they have allowed or first forbidden decays and relatively long life-times, is provided in Table \ref{UL_ShortList_Table}. These candidates also have $Q_{es}$ values that are close to zero with relatively small uncertainties, increasing the likelihood that a precise Penning trap measurement would reveal a Q value of $\approx$1 keV.

\section*{Acknowledgments}

This material is based upon work supported by the US Department of Energy, Office of Science, Office of Nuclear Physics under Award No. DE-SC0015927 and DE-FG02-93ER40789, and by the National Science Foundation under Contract No. 2111302.  
K.G.L also acknowledges support from the Gordon and Betty Moore Foundation and the Facility for Rare Isotope Beams (FRIB), which is a DOE Office of Science User Facility under Award No. DE-SC0000661.
Support was also provided by Central Michigan University.

\bibliography{Refs.bib}


\end{document}